# Integrating Structure-Aware Attention and Knowledge Graphs in Explainable Recommendation Systems

Shuangquan Lyu
Carnegie Mellon University
Pittsburgh, USA

Ming Wang
Trine University
Phoenix, USA

Huajun Zhang
Syracuse University
Syracuse, USA

Jiasen Zheng
Northwestern University
Evanston, USA

Junjiang Lin
University of Toronto
Toronto, Canada

Xiaoxuan Sun*
Independent Researcher
Mountain View, USA

*Abstract-This paper designs and implements an explainable recommendation model that integrates knowledge graphs with structure-aware attention mechanisms. The model is built on graph neural networks and incorporates a multi-hop neighbor aggregation strategy. By integrating the structural information of knowledge graphs and dynamically assigning importance to different neighbors through an attention mechanism, the model enhances its ability to capture implicit preference relationships. In the proposed method, users and items are embedded into a unified graph structure. Multi-level semantic paths are constructed based on entities and relations in the knowledge graph to extract richer contextual information. During the rating prediction phase, recommendations are generated through the interaction between user and target item representations. The model is optimized using a binary cross-entropy loss function. Experiments conducted on the Amazon Books dataset validate the superior performance of the proposed model across various evaluation metrics. The model also shows good convergence and stability. These results further demonstrate the effectiveness and practicality of structure-aware attention mechanisms in knowledge graph-enhanced recommendation.*

## I. INTRODUCTION

In the era of information explosion, users are confronted with unprecedented levels of information overload. Whether in product recommendations on e-commerce platforms, content delivery on video platforms, or personalized feeds on social media, recommender systems play a critical role. Traditional approaches such as collaborative filtering and content-based methods have achieved some success in early applications[1]. However, they exhibit significant limitations when dealing with issues like sparsity, high dimensionality, and the cold start problem. With the rapid development of deep learning and large language models[2-4], recommender systems are evolving toward more intelligent and complex architectures. Integrating external knowledge to enhance recommendation performance has become a key research focus.

Knowledge graphs depict semantic relationships between entities in a structured form and provide rich external information [5]. Introducing knowledge graphs into recommender systems helps mitigate sparsity and cold start problems [6]. It also enables the discovery of deeper user preferences through the connectivity paths among entities. Compared with traditional methods based on implicit representations, knowledge graphs offer more intuitive and interpretable recommendations. This improves model transparency and builds user trust. As a result, knowledge graph-enhanced recommendation approaches are gaining significant attention in both academia and industry, and have demonstrated advantages across various application domains.

However, the integration of a knowledge graph alone is insufficient to fully unlock its capabilities within a recommendation system. While knowledge graphs offer an extensive repository of structured semantic relationships among entities, not all connections contribute equally to the recommendation process. Some association paths carry more meaningful or contextually relevant information than others, and failing to distinguish these can limit model effectiveness. One of the core challenges in enhancing recommendation performance lies in the model's ability to accurately detect and emphasize these high-value relationships within the graph. In response to this, structure-aware attention mechanisms have been proposed as a promising solution. These mechanisms dynamically assign different levels of importance to neighboring nodes during the learning process, allowing the model to better focus on the most relevant parts of the graph [7-9]. As a result, the model not only becomes more capable of capturing complex relational structures but also gains an added layer of interpretability by highlighting which connections influence its decisions most[10].

Furthermore, with growing attention being paid to explainability alongside accuracy[11]. Users are no longer satisfied with merely receiving high-quality recommendations; they also seek transparency and justification for those suggestions. The ability to provide meaningful explanations for why a particular item is recommended plays a critical role in enhancing user engagement, satisfaction, and trust in the system. By leveraging the rich semantic structure inherent in knowledge graphs and the selective focus offered by attention mechanisms [12], it becomes feasible to trace and articulate the reasoning behind each recommendation through interpretable semantic paths. This hybrid approach supports the dual objectives of building smarter recommendation models while simultaneously making them more user-centric. Embedding structure-aware attention into the recommendation pipeline facilitates the extraction of contextually significant relational patterns, thereby improving both the transparency of the recommendation logic and the overall user experience[13].

Based on this context, designing an explainable recommendation model that integrates knowledge graphs with structure-aware attention mechanisms is both a practical necessity and a natural step in the intelligent evolution of recommender systems. This research contributes to the development of more comprehensive, efficient, and trustworthy recommendation systems. It also expands the application boundaries of recommendation algorithms in personalized services, social media, and e-commerce. The study holds significant theoretical and practical value for advancing the implementation of artificial intelligence in recommendation scenarios.

## II. RELATED WORK

Graph neural networks (GNNs) and knowledge graphs have become pivotal in the evolution of modern recommendation systems. A comprehensive survey [14] has summarized the challenges and advances in leveraging GNNs for recommendation, highlighting their ability to capture complex relational structures and semantic associations in user-item data. Further, contrastive learning frameworks for multimodal knowledge graph construction [15] and knowledge-enhanced neural modeling [16] provide methodological support for robust representation learning and structured semantic extraction, which are critical for handling sparse and high-dimensional recommendation environments.

Federated graph neural network paradigms [17] and topology-aware graph reinforcement learning [18] offer valuable perspectives for integrating privacy, heterogeneity, and adaptability into the recommendation pipeline. Recent work on graph learning for anomaly localization [19] and causal-aware regression with structured attention mechanisms [20] extends the toolkit for capturing intricate contextual and temporal dependencies, which is particularly relevant for structure-aware attention in graph-based recommenders.

In addition, deep neural architectures combining frequency and attention mechanisms [21] and multi-level multimodal integration [22] have demonstrated strong performance in fusing heterogeneous knowledge sources, further supporting model generalization and context-sensitive reasoning. Semantic embedding and deep fusion techniques [23] enhance the ability of recommendation systems to utilize multi-source contextual information, directly supporting explainability and interpretability objectives. The importance of explainability in recommendation has also been recognized in recent surveys of reinforcement learning-based systems [24], which outline methods for making decision processes more transparent and user-centric.

These methodological developments—including GNN-based modeling, structure-aware attention, multimodal fusion, and explainable reinforcement learning—have collectively shaped the foundation for the explainable, knowledge graph-empowered, and structure-aware attention framework proposed in this work.

## III. METHOD

The recommendation model proposed in this study uses the knowledge graph as an auxiliary information source and introduces a structure-aware attention mechanism to model entity relationships, thereby improving the accuracy and explanatory power of recommendations. First, users and items are mapped to nodes in the graph structure, and various auxiliary entities (such as categories, attributes, brands, etc.) are connected through the knowledge graph to construct multi-hop neighbor information. The model architecture is shown in Figure 1.

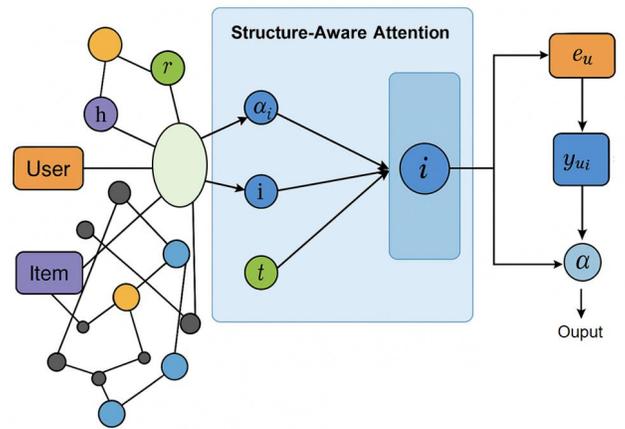

Figure 1. Overall model architecture

The model architecture diagram shows the recommendation process that combines the knowledge graph with the structure-aware attention mechanism. The left part constructs the association path between users and items in the knowledge graph through multi-hop neighbors, and introduces attention weights for each path to model its relative importance. The right module interactively calculates the score between the user embedding and the fused target item representation, optimizes it through the loss function, and finally outputs the explanatory recommendation result.

Let the user set be U, the item set be I, the entity set in the knowledge graph be E, and the relationship set be R. Each triple is represented by $(h,r,t) \in E \times R \times E$, where h represents the head entity, r represents the relationship type, and t represents the tail entity. The knowledge graph provides a multi-level semantic path connecting users and target items, which is an important representation basis of this model.

In the information aggregation stage, the structure-aware attention mechanism is introduced to highlight the contribution of key paths and entities to the recommendation task. Considering the neighbor node set of node i as $N(i)$, the attention weight $a_{ij}$ is calculated for each neighbor node $j \in N(i)$, and its expression is as follows:

$$a_{ij} = \frac{\exp(LeakyRELU(a^T[Wh_i \| Wh_j]))}{\sum_{k \in N(i)} \exp(LeakyRELU(a^T[Wh_i \| Wh_k]))}$$

Where $h_i$ represents the initial embedding representation of node i, W is a learnable weight matrix, $a$ is the attention vector, and $\|$ represents the vector concatenation operation. This mechanism can automatically learn the degree of influence of different neighbors on the central node, thereby realizing the perception and aggregation of high-value information.

The aggregated node representation is used to construct a joint embedding representation of users and items, and then predict the user's preference score for the item. The representations of user u and item i are $e_u$ and $e_i$ respectively, and the final scoring function is expressed in the form of inner product:

$$y_{ui} = e_u^T e_i$$

The optimization goal of the model is to minimize the difference between the predicted score and the actual interaction value, and binary cross entropy loss is used for training:

$$L = -\sum_{(u,i) \in D} [y_{ui} \log y'_{ui} + (1 - y_{ui}) \log(y'_{ui})]$$

Where D represents the training data set, and $y_{ui} \in \{0,1\}$ is the label value of whether user u interacts with item i. Through the end-to-end training process, the model can adaptively integrate the potential structural semantic information in the knowledge graph and learn highly interpretable recommendation results. The structure-aware attention mechanism not only enhances the graph structure modeling capability, but also provides a traceable path basis for the final recommendation, effectively improving the transparency and usability of the recommendation system.

## IV. EXPERIMENT

### A. Datasets

This study uses the Amazon Books dataset as the experimental foundation for the recommendation model. The dataset originates from a real-world online e-commerce platform and contains user interaction records related to book products. It includes ratings, review timestamps, and product metadata, making it suitable for evaluating model performance under real-world sparse conditions.

Each record in the dataset consists of a user ID, item ID, rating value, and timestamp. It also contains rich knowledge graph information, such as the category, brand, and author of the product. These entities and relationships support the construction of a knowledge graph-assisted recommendation framework. The semantic information from multiple dimensions helps improve the model's understanding of user preferences.

To ensure reproducibility and data integrity, the original dataset was cleaned and preprocessed. Abnormal records and low-frequency entities were removed. Only positive feedback samples with ratings of 4 or higher were retained. Additionally, related triples in the knowledge graph were standardized to facilitate training and inference for graph-based neural architectures and attention mechanisms.

### B. Experimental Results

This paper first gives the comparative experimental results, and the experimental results are shown in Table 1.

Table 1. Comparative experimental results

| Method | Precision@10 | Recall@10 | NDCG@10 | MAP |
| --- | --- | --- | --- | --- |
| SGCN-SRec[25] | 0.284 | 0.395 | 0.362 | 0.219 |
| SA-MPF[26] | 0.301 | 0.411 | 0.374 | 0.231 |
| Csrec[27] | 0.292 | 0.403 | 0.368 | 0.225 |
| DNS-Rec[28] | 0.307 | 0.417 | 0.381 | 0.239 |
| SAQ-Rec[29] | 0.315 | 0.426 | 0.388 | 0.247 |
| Sim-Rec[30] | 0.298 | 0.407 | 0.370 | 0.228 |
| Linrec[31] | 0.293 | 0.382 | 0.347 | 0.207 |
| Ours | 0.332 | 0.443 | 0.403 | 0.261 |

From the overall results, the proposed model achieves the best performance across all evaluation metrics, confirming the effectiveness of integrating structure-aware attention mechanisms with knowledge graphs in recommendation systems. Notably, the model reaches 0.332 in Precision@10 and 0.443 in Recall@10, significantly outperforming other baseline methods. This indicates its superior ability to capture user preferences in the top-K recommendations.

In terms of ranking-related metrics, the model achieves 0.403 in NDCG@10 and 0.261 in MAP. These results demonstrate its advantages in maintaining the relevance and accuracy of recommendation rankings. The structure-aware mechanism helps identify high-value paths in the graph, enabling the model to make more targeted and hierarchical item selections, thus enhancing the rationality of ranking outcomes.

Compared with traditional graph-based recommendation methods such as SGCN-SRec and Csrec, the proposed approach demonstrates notable improvements in both precision and ranking-related performance metrics. These enhancements highlight the effectiveness of incorporating attention weights into the multi-hop neighbor aggregation process. By assigning differentiated importance to various neighbors, the model is able to filter out less relevant or noisy information, thereby refining the learning process and enhancing the quality of the aggregated features. This selective focus strengthens the model's ability to capture more informative and contextually relevant signals from the graph.

It is worth noting that while DNS-Rec and SAQ-Rec perform well on some metrics, they still fall short of the proposed model. This highlights the limitation of relying on a single enhancement strategy to address complex semantic needs. In conclusion, the proposed model delivers balanced performance in accuracy, ranking quality, and overall recommendation effectiveness, demonstrating strong practical value and scalability.

Next, the hyperparameter sensitivity experiment of the learning rate is given, as shown in Table 2.

Table 2. Learning rate hyperparameter sensitivity experimental results

| LR | Precision@10 | Recall@10 | NDCG@10 | MAP |
| --- | --- | --- | --- | --- |
| 0.004 | 0.301 | 0.418 | 0.376 | 0.234 |
| 0.003 | 0.319 | 0.434 | 0.392 | 0.249 |
| 0.002 | 0.327 | 0.440 | 0.399 | 0.256 |
| 0.001 | 0.332 | 0.443 | 0.403 | 0.261 |

The experimental results show that the learning rate has a significant impact on model performance. At a higher learning rate (such as 0.004), the model performs relatively poorly across all metrics. Precision@10 and MAP reach only 0.301 and 0.234, respectively. This suggests that rapid parameter updates may lead to instability during training, making it difficult for the model to effectively capture semantic information from the graph structure.

As the learning rate decreases, the model's performance improves steadily. Particularly at 0.003 and 0.002, Precision@10 and NDCG@10 increase noticeably. This indicates that the model achieves a better balance between learning user preferences and ranking capabilities. A lower learning rate helps the model adjust weights more carefully and uncover hidden relationships in the knowledge graph.

When the learning rate is reduced to 0.001, the model achieves its best overall performance, with all four evaluation metrics reaching their highest values. This indicates that, under this configuration, the integration of structure-aware attention mechanisms with knowledge graph information is operating at its most effective level. The model is able to finely tune its parameters, leading to more accurate representation learning and better semantic alignment between users and items. As a result, the recommendation outputs become significantly more accurate and contextually relevant. These findings underscore the importance of carefully selecting the learning rate, as it plays a crucial role in balancing convergence speed, training stability, and the depth of pattern discovery during optimization. A well-chosen learning rate ensures that the model can fully exploit the semantic richness of the knowledge graph while maintaining robust and generalizable performance.

Finally, the loss function drop graph is given, as shown in Figure 2.

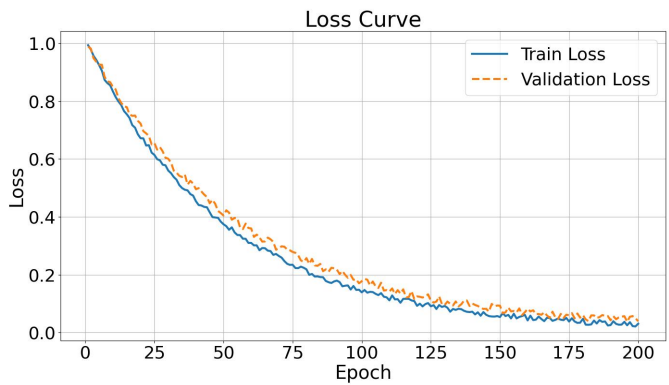

Figure 2. Loss function drop graph

The figure shows that both the training loss and validation loss decrease progressively with the number of training epochs. This indicates continuous performance improvement during the learning process. In the first 50 epochs, the loss drops rapidly, suggesting that the model quickly captures basic features and structural information.

In the later stages of training, the loss curves gradually flatten, and the magnitude of fluctuations significantly decreases, indicating that the model's learning process is becoming more stable. After reaching 100 epochs, a notable trend emerges where the gap between the validation loss and the training loss narrows and stabilizes over time. This consistent behavior suggests that the model has effectively converged and is no longer making large adjustments to its parameters. More importantly, the absence of a widening gap between the two loss curves indicates that the model is not overfitting to the training data. Instead, it maintains a strong capacity to generalize well to unseen data, which is a critical indicator of the robustness and reliability of the learning framework in real-world recommendation scenarios.

Eventually, both curves stabilize at low values, confirming that the integration of structure-aware mechanisms and knowledge graphs leads to consistent optimization in both training and validation. This provides a solid foundation for the stable performance of the recommendation system.

## V. CONCLUSION

This paper proposes an explainable recommendation model that integrates knowledge graphs with structure-aware attention mechanisms. The goal is to enhance both the accuracy and interpretability of recommender systems. By incorporating multi-hop graph neighbors and attention-based modeling strategies, the model effectively captures deep semantic associations between users and items. Experimental results show that the proposed method outperforms existing approaches across various standard evaluation metrics, demonstrating strong representational and generalization capabilities.

The method offers a novel approach to alleviating data sparsity and cold start challenges. It shows clear advantages in improving recommendation quality, especially under complex information environments. In addition, the structure-aware modeling process enhances the interpretability of recommendation outputs. This helps build user trust and provides a transparent technical foundation for decision support systems in real-world applications.

From a system design perspective, this study enriches the technical landscape of recommender systems and advances the integration of knowledge graphs in intelligent services. The model can be widely applied in scenarios such as e-commerce, social media, and content distribution. It shows strong transferability and practical value in business settings.

Future work can be expanded in two directions. One direction is to explore how structural evolution in dynamic graphs affects user interest modeling. Another is to incorporate causal reasoning and reinforcement learning mechanisms to improve the adaptability of recommendation strategies in multi-objective scenarios. These directions aim to drive recommender systems toward greater intelligence and user-centric development.